\documentclass[10pt,superscriptaddress,nofootinbib,notitlepage,tightenlines,twocolumn, pra]{revtex4-2}
\usepackage{hyperref}
\usepackage{verbatim}
\usepackage{amsmath}
\usepackage{latexsym}
\usepackage{revsymb}
\usepackage{yfonts}
\usepackage{ifthen}

\usepackage{natbib}
\usepackage{amsfonts}
\usepackage{amsmath}
\usepackage{amssymb}
\usepackage{amsthm}
\usepackage{graphicx}
\usepackage{bm}
\usepackage{bbm}
\usepackage{epsfig,color,amssymb}
\usepackage{subfigure}
\usepackage{amsfonts}
\usepackage{amscd}
\usepackage{amsmath}
\usepackage{multirow}
\usepackage{chemarrow}
\usepackage{dcolumn}
\usepackage{bm}
\usepackage{graphicx}
\usepackage{enumerate}
\usepackage{epsfig}
\usepackage{subfigure}
\usepackage{xcolor}
\usepackage{multirow}
\usepackage{braket}
\usepackage{comment}
\usepackage{enumitem}
\usepackage{amsthm}
\renewcommand{\emph}[1]{\textit{#1}}

\begin{document}
	
\title{Long distance twin-field quantum key distribution with entangled sources}
\author{Bing-Hong Li}
\author{Yuan-Mei Xie}
\author{Zhao Li}
\author{Chen-Xun Weng}
\author{Chen-Long Li}
\author{Hua-Lei Yin}\email{hlyin@nju.edu.cn}
\author{Zeng-Bing Chen}\email{zbchen@nju.edu.cn}
\affiliation{National Laboratory of Solid State Microstructures, School of Physics and Collaborative Innovation Center of Advanced Microstructures, Nanjing University, Nanjing 210093, China}


\begin{abstract}
Twin-field quantum key distribution (TFQKD), using single-photon-type interference, offers a way to exceed the rate-distance limit without quantum repeaters. However, it still suffers from the photon losses and dark counts, which impose an ultimate limit on its transmission distance. In this letter, we propose a  scheme to implement TFQKD with an entangled coherent state source in the middle to increase its range, as well as comparing its performance under coherent attacks with that of TFQKD variants. Simulations show that our protocol has a theoretical distance advantage of 400 kilometers. Moreover, the scheme has great robustness against the misalignment error and finite-size effects. Our work is a promising step toward long-distance secure communication and is greatly compatible with future global quantum network.
\end{abstract}


\maketitle

Quantum key distribution (QKD)~\cite{bennett2014quantum}, by utilizing the fundamental principles of quantum mechanics, enables two distant users to share a secret key in the presence of an eavesdropper. Due to its great potential in cryptography, in the past three decades QKD has significantly developed both theoretically and experimentally~\cite{lo2012measurement,wang2013three,zhou2016making} and has been successfully implemented over long distance~\cite{PhysRevLett.117.190501,PhysRevLett.121.190502}. However, there still exist some challenges to be overcome in the implementation of QKD, a formidable one of which is how to perform QKD at longer distance. In the literature, without quantum repeaters, the key rate of QKD scales linearly with the channel transmittance~\cite{takeoka2014fundamental,pirandola2017fundamental} $\eta$. To be specific, there is a fundamental limit called Pirandola–Laurenza–Ottaviani–Banchi (PLOB) bound~\cite{pirandola2017fundamental},
which demonstrates that the secret key rate without quantum repeaters must satisfy $R \le -\log_{2}{(1-\eta)} $.

Recently, a breakthrough called twin-field QKD~\cite{lucamarini2018overcoming} (TFQKD) has been put forward to overcome the PLOB bound. By exploiting the single-photon-type interference in the untrusted relay, TFQKD can provide a secret key rate proportional to the square-root of channel transmittance, thus allowing unprecedented distance coverage.
Many variants of TFQKD protocols have been proposed~\cite{ma2018phase,tamaki2018information,wang2018twin,hua2019measurement,cui2019twin,curty2019simple,lin2018simple,primaatmaja2019versatile,yin2019coherent,maeda2019repeaterless,Zhang:19,Xu:20,xu2020sending} and some protocols have been demonstrated experimentally~\cite{minder2019experimental,PhysRevLett.123.100505,PhysRevX.9.021046,zhong2019proof,fang2020implementation,PhysRevLett.124.070501}. Remarkably, phase-matching QKD~\cite{ma2018phase} (PM-QKD) and  sending-or-not-sending TFQKD~\cite{wang2018twin} (SNS-TFQKD) have achieved over 500 km.
Moreover, the idea of TFQKD has been implemented to other quantum tasks, such as device independent QKD~\cite{Xie:21}, quantum conference key agreement~\cite{zhao2020phase,cao2021coherent} and quantum secret sharing~\cite{Gu:21,gu2021secure}.
Nonetheless, just like other protocols, TFQKD is also plagued by the dark counts and photon losses. Dark counts will induce an intrinsic error because it is difficult to discriminate which click of the detector is caused by dark counts. As the transmission distance becomes longer, photon losses will also be larger, enhancing the influence of dark count. Thus at long distance, the secret key rate will drop dramatically, which restricts the maximum transmission distance of TFQKD.

Enlightened by the ideas of MDI-QKD with entangled photon sources in the middle~\cite{xu2013long}, here we propose a  protocol based on entangled coherent state (ECS) sources to further improve the transmission distance of TFQKD.
ECS is an important quantum resource in quantum information processing~\cite{PhysRevA.45.6811,RevModPhys.85.1103} and has become of widespread interest in quantum teleportation ~\cite{PhysRevA.64.052308,PhysRevLett.91.017902},  Bell-like inequality violation~\cite{PhysRevA.80.022111},  quantum repeater~\cite{PhysRevLett.105.160501} and quantum metrology~\cite{joo2011quantum}. Although it is a challenge to generate ECS with high fidelity and frequency, there are numerous works pursuing it.
Methods to generate ECS have been proposed based on nonlinear Kerr evolution~\cite{PhysRevLett.91.017902}, electromagnetically induced transparency~\cite{PhysRevA.67.023811} and cavity quantum electrodynamics system~\cite{liu2016generation}. Importantly, the optical generation of ECS has been experimentally demonstrated~\cite{Israel:19,ourjoumtsev2009preparation}.

In our protocol, Charlie, in the middle between Alice and Bob, holds the ECS source while Alice and Bob each prepare the coherent states in the $Z$ basis or cat states in the $X$ basis, randomly and independently.
Between Alice (Bob) and Charlie is an untrusted relay David (Fred). After the preparation of states, Alice (Bob) and Charlie send their pulses to  David (Fred). Then David and Fred perform an ECS measurement to achieve the interference. Different from other protocols of TFQKD, in our protocol a single photon just goes through a quarter of the total distance between Alice and Bob. Although the requirement of coincident detection of David and Fred leads to the key rate still scaling with half of the total distance, our protocol can further extend the distance of TFQKD by reducing the signal to noise ratio at the measurement nodes. Hence it provides a promising candidate for implementing long-distance QKD experimentally.

\begin{figure}[t]
	\centering
	\centering\includegraphics[width=\linewidth]{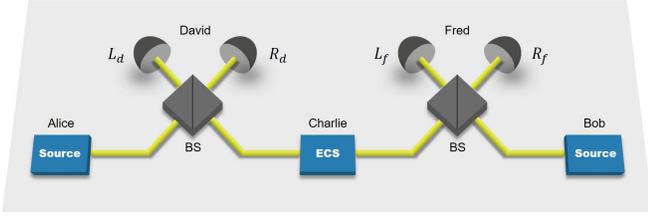}
	\caption{Alice (Bob) randomly prepares the coherent states in the $Z$ basis or cat states in the $X$ basis and sends her (his) pulses to the untrusted relay David (Fred). Charlie prepares the ECS and sends one pulse to David and another to Fred. David and Fred perform an ECS measurement on the pulses they receive, respectively.}
	\label{fig:1}
\end{figure}

Our protocol is detailed below and the devices are shown in Fig.~\ref{fig:1}:

\noindent\textbf{Step 1.}——Alice and Bob each randomly prepare the coherent states $\left|\pm \alpha\right\rangle$ in the $Z$ basis or the cat states $\left|\xi^\pm (\alpha)\right\rangle=(\left|\alpha\right\rangle \pm \left|-\alpha\right\rangle)/\sqrt{N_\pm}$ in the $X$ basis, while untrusted source Charlie prepares the ECS source $\left | \Phi^-  \right\rangle=(\left|\alpha\right\rangle\left|\alpha\right\rangle-\left|-\alpha\right\rangle\left|-\alpha\right\rangle)/\sqrt{2(1-e^{-4\mu})}$, where $\left|\pm \alpha\right\rangle$ are the coherent states with opposite phases, $N_\pm=2(1\pm e^{-2\mu})$ are normalization constants and $\mu=|\alpha|^2$ is the intensity of the coherent state.
The probability of choosing the $Z (X)$ basis is denoted as $p_z (p_x)$.
Then Alice and Bob record their binary choices of coherent states and label them by $a_i,~b_j\in\{0,~1\}$ for encoding.

\noindent\textbf{Step 2. }——Alice, Bob and Charlie send the optical pulses  to the two untrusted relays, David and Fred, each of whom holds a beam splitter (BS) and two threshold detectors $L_d$ ($L_f$) and $R_d$ ($R_f$). Then David and Fred preform an ECS measurement on the two received pulses and publicly broadcast their detection results to Alice and Bob. A successful detection is defined as the case where  one and only one of two detectors clicks.

\begin{table}[b]
	\centering
	\caption{Alice and Bob post-select the successful measurement results when they use the same basis. Parameters $L_{d(f)}$ = 0 and 1 represent that the detector $L$ of David (Fred) does not click and clicks, respectively; parameters $R_{d(f)}$ = 0 and 1 represent that the detector $R$ of David (Fred) does not click and clicks, respectively.}
    \begin{tabular}{c|cccc}
        \hline
		&&$L_{d}R_{d}L_{f}R_{f}$\\
		\hline
		basis    & 1010  &1001  &0110  &0101 \\
		\hline
		Z basis & no flip &  flip & flip & no flip\\
		X basis    & flip  &flip & flip & flip \\
		\hline
    \end{tabular}
\label{table1}
\end{table}

\noindent\textbf{Step 3. }——Alice and Bob announce their choices of bases
over an authenticated classical channel. They only keep the data of the same basis and discard the rest. Then Bob decides whether to flip his key bit $b_j$ according to different measurement results and basis selections.
The operation of Bob's key bits according to the successful detection is shown in Table \ref{table1}. For the $Z$ basis, Bob flips his key bit if David and Fred announce that $L\ (R)$ and $R\ (L)$ detector clicks. For the $X$ basis, Bob always flips his key bit.
A part of the data generated in the $Z$ basis are used for error correction and others are used for key generation, while those generated in the $X$ basis are used for parameter estimation.

\noindent\textbf{Step 4. }——Alice and Bob repeat Steps 1-3 sufficient times to generate a string of raw data. Then they perform parameter estimation, error correction and privacy amplification to extract the final secure keys.

The key idea of our scheme is to add an additional ECS source owned by Charlie and two signals announced by David and Fred, which indicates whether the entangled state is successfully shared between Alice and Bob.
According to the signals from David and Fred, we can successfully select the correct data and exclude the failed entanglement distribution events upfront from being used in the trial.

The security proof of our prepare-and-measure protocol is based on the virtual entanglement-based protocol.
The coherent states $\left|\pm \alpha\right\rangle$ and the cat states $\left|\xi^\pm (\alpha)\right\rangle$ preparation process in the protocol can be purified into an entanglement-based case, where Alice and Bob randomly exploit the $Z$ and $X$ basis to act a local measurement on the  system of the entangled state $\left | \psi  \right\rangle=(\left|z\right\rangle\left|\alpha\right\rangle+ \left|-z\right\rangle\left|-\alpha\right\rangle)/\sqrt{2}$. Moreover, the local measurement can be delayed until David and Fred announce their ECS measurement results, so that we can get a virtual entanglement-based protocol.
In the asymptotic limit, we can assume $p_x=0$, so the secret key rate of our protocol is~\cite{yin2019coherent,bennett1992quantum}
	$R=Q_Z[1-fh(E_Z)-h(E_X)]$,
where $Q_Z$ is the gain in the $Z$ basis, $h(x)=-x\log_{2}{x}-(1-x)\log_{2}{(1-x)}$ is the Shannon entropy, $f=1.1$ is the error correction inefficiency and $E_Z$ ($E_X$) is the bit error rate of the $Z (X)$ basis.

\begin{figure}[t]
	\centering
	\centering\includegraphics[width=8.6cm,height=6cm]{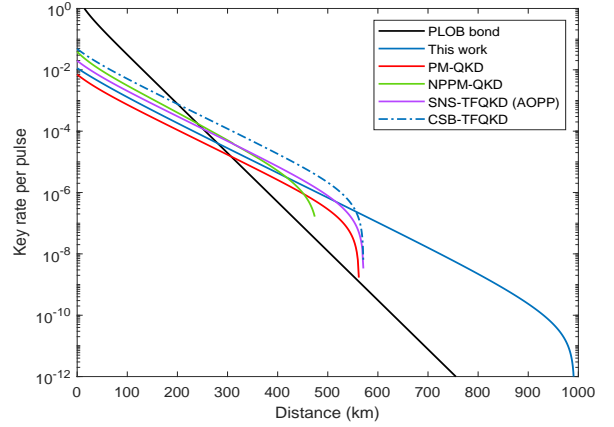}
	\caption{The key rate of our protocol compared with the variants of TFQKD. We set the dark count rate $p_d=10^{-7}$ and misalignment error of the Z basis $e_{d}=0.03$. The intensity of the coherent state for each transmission distance is optimized.}
	\label{fig:2}
\end{figure}
\begin{figure}[t]
	\centering
	\includegraphics[width=8.6cm,height=6cm]{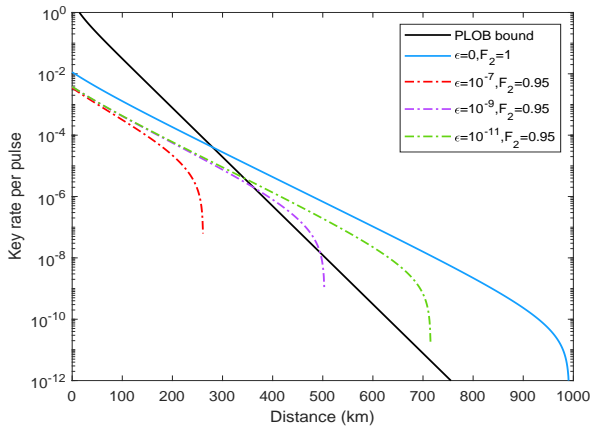}
	\caption{The key rate of our protocol with different fidelity of ECS and cat states. The intensity of the coherent state for each transmission distance is optimized.}
	\label{fig:3}
\end{figure}

Considering the practical case with threshold detectors and lossy channels which are supposed to be symmetrical for Alice and Bob. The distances of Alice-David, David-Charlie, Charlie-Fred and Fred-Bob are all $L/4$.
The dark count rate of detector is set as $p_d=10^{-7}$.  $\eta_d=85\%$ is the  efficiency of detector and $\beta=0.16$~dB/km stands for the inherent loss of ultralow-loss fiber.
Define the detection probability of identical (different) raw key bits shared between Alice and Bob in the $X$ and $Z$ bases as $Q^C_X~(Q^E_X)$ and $Q^C_Z~(Q^E_Z)$, respectively.
The misalignment error of the $X$ basis is set as $e_{d_X}=0$ since Bob always flips his key bit in the $X$ basis, which leads to the bit error rate $E_X$  not sensitive to interference. Thus, the bit error rate of the $X$ basis $E_X=Q^E_X/(Q^E_X+Q^C_X)$, while the bit error rate of the $Z$ basis $E_Z=[e_dQ^C_Z+(1-e_d)Q^E_Z]/(Q^E_Z+Q^C_Z)$. Here we use $e_d$ to represent the misalignment error of the  $Z$ basis for simplicity.

As shown in Fig.~\ref{fig:2}, we compare the key rate of our protocol with PM-QKD~\cite{ma2018phase}, no-phase-post-matching-QKD~\cite{cui2019twin,curty2019simple} (NPPM-QKD), SNS-TFQKD using active odd-party paring (AOPP) method~\cite{xu2020sending}, coherent state based-TFQKD~(CSB-TFQKD)~\cite{yin2019coherent} and the PLOB bound.
We present detailed simulation in supplement 1.
It turns out that our protocol can exceed the PLOB bond when  $L>250$ km. Although at short distance the key rate of our protocol is not superior, at long distance our protocol outperforms any other protocols and has a maximum transmission distance of approximately 1000 km.
Even at the distance of 950 km, our protocol can still have a secret key rate of $10^{-10}$ bits per pulse, which may be enough for practical quantum communication when the signal generator becomes more efficient.

\begin{figure}[t]
	\centering
	\includegraphics[width=8.6cm,height=6cm]{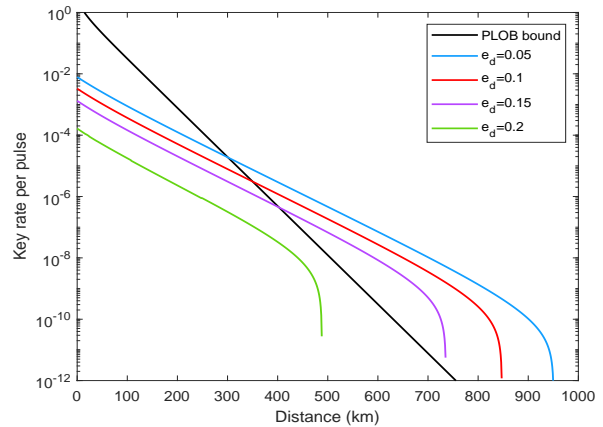}
	\caption{The key rate of our protocol under different misalignment error of Z basis, with the dark count rate $p_d=10^{-7}$. We optimize the intensity of coherent state for each transmission distance and each misalignment error.}
	\label{fig:4}
\end{figure}

Considering the fact that the generation of ECS and cat states is still difficult and in practical cases the fidelity of them is usually not 1~\cite{hacker2019deterministic,glancy2008methods}, we analyze the performance of our protocol under this non-ideal scenario.
Note that in this case the density matrix of the Z and X bases are not strictly identical, and thus the phase error rate of the Z basis $E^{\rm ph}_{Z}$ is not equal to $E_X$. To estimate the bound of $E^{\rm ph}_{Z}$, we introduce the imbalance of the quantum coin $\Delta$~\cite{lo2007security, koashi2009simple}.
Define $F_1=1-\epsilon$ as the fidelity  between  cat states and coherent states generated by Alice (Bob) and $F_2$ as the fidelity of ECS generated by Charlie. Fig.~\ref{fig:3} shows the performance of our protocol under different $F_1$, i.e. $\epsilon$, and $F_2$.
The result shows that when $\epsilon=10^{-9}$ and $F_2=95\%$, our protocol can still surpass the PLOB bound.
Detailed analysis of security and simulation can be found in supplement 1.

In the following, we draw the variation of the key generation rates with different optical misalignment errors of the $Z$ basis ($e_{d}$), as shown in Fig.~\ref{fig:4}.  The result shows that our scheme can still break the PLOB bond even if the misalignment error is as large as $15\%$,  indicating  great robustness of our protocol against environment noise.

\begin{figure}[t]
	\centering
	\includegraphics[width=8.6cm,height=6cm]{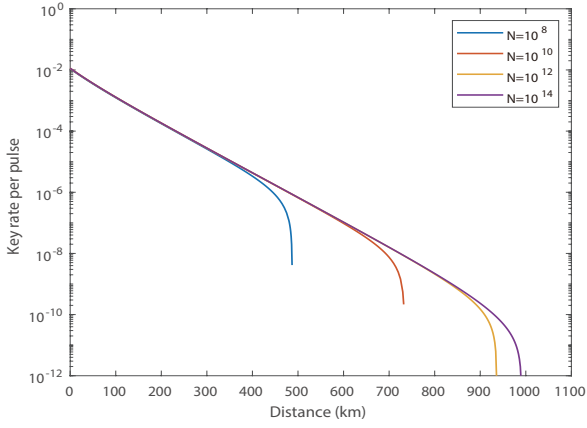}
	\caption{The finite-key rate of our protocol with different data sizes of $N=10^{8}, 10^{10}, 10^{12}$ and $10^{14}$. The security bound and correct bound are set to be $\epsilon_{\rm sec} = 10^{-10}$ and $\epsilon_{\rm cor} = 10^{-15}$. The intensity of the coherent state and the probability of choosing the  $Z$ basis for each transmission distance are optimized.}
	\label{fig:5}
\end{figure}

In the end, we analyse the performance of our protocol under the finite-size effects.
Considering the practical case with limited amount of data shared between Alice and Bob, the effect of statistical fluctuation cannot be neglected, causing the estimated phase error rate of the  $Z$ basis to be increased.
Besides, $p_x$ will no longer be treated as zero, which will introduce an extra loss of the key capacity.
Here we use the techniques of finite-key analysis~\cite{yin2019finite}, simulating the key rate with the secret bound $\epsilon_{\rm sec} = 10^{-10}$ and the correct bound $\epsilon_{\rm cor} = 10^{-15}$ under different data sizes $N$.
The  key rate $R$ satisfies
\begin{equation}
	R=\frac{1}{N}\left\{n^Z[1-h(\bar{E}_{Z}^{\rm ph})]-\lambda_{\rm EC}-\log_{2}{\frac{8}{\epsilon_{\rm cor}\epsilon_{\rm sec}^2 }}\right\},
\end{equation}
where $n^Z = (1-p_x)^2 N$ is the key length of the  $Z$ basis,  $\bar{E}_{Z}^{\rm ph}$ is the upper bound of the phase error rate of the  $Z$ basis estimated by the bit error rate of the  $X$ basis, $\lambda_{\rm EC} = fn^Zh(E_Z)$ is the error correction leakage.
As shown in Fig.~\ref{fig:5}, when the data size is $10^{14}$, the performance of our protocol is closely near to that in the  asymptotic limit.  Even the data size is as small as $N=10^{10}$, the transmission distance can still reach 700 km.

In summary, we propose a  TFQKD scheme which introduces an  untrusted entangled photon source and two untrusted relays in the middle.
By successfully sharing  ECSs between two distant users through two relays, our protocol can overcome the PLOB bound and achieve a distance of nearly 1000 km. Moreover, it shows great robustness against environment noise and finite-size effects. Although it is a challenge to utilize ECS and cat states encoding in QKD experiment with high efficiency, we could expect this issue to be solved in the near future.
Our method is highly compatible with the future quantum networks and
paves the ground for the application of long-distance quantum communication.

\noindent\textbf{Acknowledgments}\\
We gratefully acknowledge support from the National Natural Science Foundation of China (under Grant No. 61801420); the Key-Area Research and Development Program of Guangdong Province (under Grant No. 2020B0303040001); the Fundamental Research Funds for the Central Universities (under Grant No. 020414380182); the  Natural Science Foundation of Jiangsu Province (under Grant No. BK20211145)

\noindent\textbf{Author Contributions}\\
B-H Li and Y-M Xie contributed equally to the work presented in this paper.

\begin{widetext}
\appendix








\section{Detailed Theoretical Model and Simulation}
In this section we present detailed theoretical model on how to calculate the key rate.
The detection operation of the threshold detectors can be characterized by two measurement operators as follows
\begin{equation}
	\begin{aligned}
		F^c &= \sum_{n=0}^{\infty}[1-(1-p_d)(1-\eta_d)^n]
		\left|n \right \rangle  \left \langle n \right|,\\
		F^{nc} &=I-F^c= \sum_{n=0}^{\infty}(1-p_d)(1-\eta_d)^n
		\left|n \right \rangle  \left \langle n \right|,
	\end{aligned}
\end{equation}
where $F^c$ denotes click and $F^{nc}$ denotes no click; $I$ is the identity operator, $p_d$  and $\eta_d$ are the dark count rate and efficiency of the detector, respectively.
To simplify the calculation we can combined the transmittance of channel and the efficiency of threshold detector, which will not affect the calculation as has been widely applied in QKD simulation. In our protocol we just consider the symmetric channel. Thus, the two measurement operators of equivalent threshold detector are
\begin{equation}
	\begin{aligned}
		F^c &= \sum_{n=0}^{\infty}[1-(1-p_d)(1-\eta)^n]
		\left|n \right \rangle  \left \langle n \right|,\\
		F^{nc} &=I-F^c= \sum_{n=0}^{\infty}(1-p_d)(1-\eta)^n
		\left|n \right \rangle  \left \langle n \right|,
	\end{aligned}
\end{equation}
where $\eta=\eta_d \eta_t$ is the total efficiency, $\eta_t$ is the transmittennce of channel.

For the case of the Z basis in our protocol with an ECS source $\left|\Phi^- \right \rangle$ in the middle, the evolution of quantum states after passing through BS is
\begin{equation}
	\begin{aligned}
		\left|\alpha \right \rangle_a \left|\Phi^- \right \rangle_{c} \left|\alpha \right \rangle_b &\overset{BS}{\rightarrow} \frac{1}{N^-}(\left|\sigma \right \rangle_1 \left|0 \right \rangle_2 \left|\sigma \right \rangle_3 \left|0 \right \rangle_4 - \left|0 \right \rangle_1 \left|\sigma \right \rangle_2 \left|0 \right \rangle_3 \left|-\sigma \right \rangle_4)
		:=\left|\phi_Z^{0,0} \right \rangle_{1234}, \\
		\left|\alpha \right \rangle_a \left|\Phi^- \right \rangle_{c} \left|-\alpha \right \rangle_b &\overset{BS}{\rightarrow}
		\frac{1}{N^-}(\left|\sigma \right \rangle_1 \left|0 \right \rangle_2 \left|0 \right \rangle_3 \left|\sigma \right \rangle_4
		- \left|0 \right \rangle_1 \left|\sigma \right \rangle_2 \left|-\sigma \right \rangle_3 \left|0 \right \rangle_4)
		:=\left|\phi_Z^{0,1} \right \rangle_{1234}, \\
		\left|-\alpha \right \rangle_a \left|\Phi^- \right \rangle_{c} \left|\alpha \right \rangle_b &\overset{BS}{\rightarrow}
		\frac{1}{N^-}(\left|0 \right \rangle_1 \left|-\sigma \right \rangle_2 \left|\sigma \right \rangle_3 \left|0 \right \rangle_4
		- \left|-\sigma \right \rangle_1 \left|0 \right \rangle_2 \left|0 \right \rangle_3 \left|-\sigma \right \rangle_4)
		:=\left|\phi_Z^{1,0} \right \rangle_{1234}, \\
		\left|-\alpha \right \rangle_a \left|\Phi^- \right \rangle_{c} \left|-\alpha \right \rangle_b &\overset{BS}{\rightarrow}
		\frac{1}{N^-}(\left|0 \right \rangle_1 \left|-\sigma \right \rangle_2 \left|0 \right \rangle_3 \left|\sigma \right \rangle_4
		- \left|-\sigma \right \rangle_1 \left|0 \right \rangle_2 \left|-\sigma \right \rangle_3 \left|0 \right \rangle_4)
		:=\left|\phi_Z^{1,1} \right \rangle_{1234},
	\end{aligned}
\end{equation}
where $\sigma$ := $\sqrt2 \alpha$ and the subscript $a, b, c$ denote Alice, Bob and Charlie. For simplicity we use subscript $1, 2, 3, 4$ to denote the detecor $L_d$, $R_d$, $L_f$ and $R_f$, respectively. Then we can get the correct gain and the error gain of the Z basis as follows
\begin{equation}
	\begin{split}
		Q_Z^C=&\frac{1}{4}Tr  \{ \ (F^c_1 F^{nc}_2 F^c_3 F^{nc}_4 + F^{nc}_1 F^c_2 F^{nc}_3 F^c_4  )  [P(|\phi_Z^{0,0}  \rangle_{1234})+P(|\phi_Z^{1,1}  \rangle_{1234})  ]  \\
		&+(F^c_1 F^{nc}_2 F^{nc}_3 F^c_4 + F^{nc}_1 F^c_2 F^c_3 F^{nc}_4)[P(|\phi_Z^{0,1}  \rangle_{1234})+P(|\phi_Z^{1,0}  \rangle_{1234})]  \} \\
		=&\frac{2}{N_-}\{(1-p_d)^2[1-(1-p_d)e^{-2\mu\eta}]^2+p_d^2(1-p_d)^2(e^{-4\mu\eta}-2e^{-4\mu})\}, \\
		Q_Z^E=&\frac{1}{4}Tr\{(F^c_1 F^{nc}_2 F^{nc}_3 F^c_4 + F^{nc}_1 F^c_2 F^c_3 F^{nc}_4)[P(|\phi_Z^{0,0}  \rangle_{1234})+P(|\phi_Z^{1,1}  \rangle_{1234})]\\
		&+(F^c_1 F^{nc}_2 F^c_3 F^{nc}_4 + F^{nc}_1 F^c_2 F^{nc}_3 F^c_4)[P(|\phi_Z^{0,1}  \rangle_{1234})+P(|\phi_Z^{1,0} \rangle_{1234})]
		\} \\
		=&\frac{4}{N_-}\{p_d(1-p_d)^2e^{-2\mu\eta}[1-(1-p_d)e^{-2\mu\eta}]-p_d^2(1-p_d)^2e^{-4\mu}\},
	\end{split}
\end{equation}
where $P(\left|\phi \right \rangle)= \left|\phi \right \rangle \left \langle \phi \right|$, $\eta=\eta_d \times 10^{-\beta L/40}$, $\beta$ is the intrinsic loss of fiber channel and $L$ is the distance between Alice and Bob.
Similarly, we can also calculate the correct gain and error gain of the X basis
\begin{equation}
	\begin{split}
		Q_X^C
		=&\frac{1}{N_-}\{(1-p_d)^2[1-(1-p_d)e^{-2\mu\eta}][1-(1-3p_d)e^{-2\mu\eta}]
		+p_d^2(1-p_d)^2(e^{-4\mu\eta}+e^{-8\mu+4\mu\eta})\\
		&+(1-p_d)^2[e^{-4\mu}-(1-p_d)e^{-4\mu+2\mu\eta}][e^{-4\mu}-(1-3p_d)e^{-4\mu+2\mu\eta}]
		-8p_d^2(1-p_d)^2e^{-4\mu}\}, \\
		Q_X^E
		=&\frac{1}{N_-}\{(1-p_d)^2[1-(1-p_d)e^{-2\mu\eta}][1-(1-3p_d)e^{-2\mu\eta}]
		+p_d^2(1-p_d)^2(e^{-4\mu\eta}-e^{-8\mu+4\mu\eta})\\
		&-(1-p_d)^2[e^{-4\mu}-(1-p_d)e^{-4\mu+2\mu\eta}][e^{-4\mu}-(1-3p_d)e^{-4\mu+2\mu\eta}]\}. \\
	\end{split}
\end{equation}

\section{Security Analysis}

In this section we give the security analysis when Alice, Bob and Charlie generate their states imperfectly.
First, the entangled coherent states (ECS) $\left | \Phi^-  \right\rangle$ generated by Charlie may be non-ideal. We assume the actual density matrix generated by Charlie as follows
\begin{equation}
	\begin{aligned}
		\rho_{\left | \Phi^-  \right\rangle}=&F_2 \left | \Phi^-  \right\rangle \left \langle \Phi^-  \right | + (1-F_2)/3 \left | \Phi^+  \right\rangle \left \langle \Phi^+  \right | \\
		&+ (1-F_2)/3 \left | \Psi^-  \right\rangle \left \langle \Psi^-  \right |
		+ (1-F_2)/3 \left | \Psi^+  \right\rangle \left \langle \Psi^+  \right |,
	\end{aligned}
\end{equation}
where $\left | \Phi^\pm  \right\rangle=(\left|\alpha\right\rangle\left|\alpha\right\rangle \pm \left|-\alpha\right\rangle\left|-\alpha\right\rangle)/\sqrt{2(1\pm e^{-4\mu})}$ and $\left | \Psi^\pm  \right\rangle=(\left|\alpha\right\rangle\left|-\alpha\right\rangle \pm \left|-\alpha\right\rangle\left|\alpha\right\rangle)/\sqrt{2(1\pm e^{-4\mu})}$ are four ECSs, $F_2$ denotes the fidelity of ECS.	
Thus, we introduce the cases of Charlie produces ECSs $\ket{\Phi^+}$, $\ket{\Psi^-}$ and $\ket{\Psi^+}$, respectively.

\subsection{Charlie produces $\ket{\Phi^+}$}
In this case the correct and error gain of the Z and X bases are:
\begin{equation}
	\begin{split}
		Q_Z^C(\ket{\Phi^+})
		=&\frac{2}{N_+}\{(1-p_d)^2[1-(1-p_d)e^{-2\mu\eta}]^2+p_d^2(1-p_d)^2(e^{-4\mu\eta}+2e^{-4\mu})\}, \\
		Q_Z^E(\ket{\Phi^+})
		=&\frac{4}{N_+}\{p_d(1-p_d)^2e^{-2\mu\eta}[1-(1-p_d)e^{-2\mu\eta}]+p_d^2(1-p_d)^2e^{-4\mu}\}, \\
		Q_X^C(\ket{\Phi^+})
		=&\frac{1}{N_+}\{(1-p_d)^2[1-(1-p_d)e^{-2\mu\eta}][1-(1-3p_d)e^{-2\mu\eta}]
		+p_d^2(1-p_d)^2(e^{-4\mu\eta}-e^{-8\mu+4\mu\eta})\\
		&-(1-p_d)^2[e^{-4\mu}-(1-p_d)e^{-4\mu+2\mu\eta}][e^{-4\mu}-(1-3p_d)e^{-4\mu+2\mu\eta}], \\
		Q_X^E(\ket{\Phi^+})
		=&\frac{1}{N_+}\{(1-p_d)^2[1-(1-p_d)e^{-2\mu\eta}][1-(1-3p_d)e^{-2\mu\eta}]
		+p_d^2(1-p_d)^2(e^{-4\mu\eta}+e^{-8\mu+4\mu\eta})\\
		&+(1-p_d)^2[e^{-4\mu}-(1-p_d)e^{-4\mu+2\mu\eta}][e^{-4\mu}-(1-3p_d)e^{-4\mu+2\mu\eta}]+8p_d^2(1-p_d)^2e^{-4\mu}\}. \\
	\end{split}
\end{equation}
Then we have
\begin{equation}
	\begin{aligned}
		E_Z(\ket{\Phi^+})=&[e_d Q_Z^C(\ket{\Phi^+}) + (1-e_d) Q_Z^E(\ket{\Phi^+})]/(Q_Z^C(\ket{\Phi^+})+Q_Z^E(\ket{\Phi^+})),\\
		E_X(\ket{\Phi^+})=&Q_X^E(\ket{\Phi^+})/(Q_X^C(\ket{\Phi^+})+Q_X^E(\ket{\Phi^+}),
	\end{aligned}
\end{equation}
where $E_Z(\ket{X})$ and $E_X(\ket{X})$ represent bit error rate of the Z and X bases when Charlie prepares $\ket{X}$.

\subsection{Charlie produces $\ket{\Psi^-}$}
In this case the correct and error gain of the Z and X bases are:
\begin{equation}
	\begin{split}
		Q_Z^C(\ket{\Psi^-})
		=&\frac{4}{N_-}\{p_d(1-p_d)^2e^{-2\mu\eta}[1-(1-p_d)e^{-2\mu\eta}]-p_d^2(1-p_d)^2e^{-4\mu}\}, \\
		Q_Z^E(\ket{\Psi^-})
		=&\frac{2}{N_-}\{(1-p_d)^2[1-(1-p_d)e^{-2\mu\eta}]^2+p_d^2(1-p_d)^2(e^{-4\mu\eta}-2e^{-4\mu})\}, \\
		Q_X^C(\ket{\Psi^-})
		=&\frac{1}{N_-}\{(1-p_d)^2[1-(1-p_d)e^{-2\mu\eta}][1-(1-3p_d)e^{-2\mu\eta}]
		+p_d^2(1-p_d)^2(e^{-4\mu\eta}-e^{-8\mu+4\mu\eta})\\
		&-(1-p_d)^2[e^{-4\mu}-(1-p_d)e^{-4\mu+2\mu\eta}][e^{-4\mu}-(1-3p_d)e^{-4\mu+2\mu\eta}]\}, \\
		Q_X^E(\ket{\Psi^-})
		=&\frac{1}{N_-}\{(1-p_d)^2[1-(1-p_d)e^{-2\mu\eta}][1-(1-3p_d)e^{-2\mu\eta}]
		+p_d^2(1-p_d)^2(e^{-4\mu\eta}+e^{-8\mu+4\mu\eta})\\
		&+(1-p_d)^2[e^{-4\mu}-(1-p_d)e^{-4\mu+2\mu\eta}][e^{-4\mu}-(1-3p_d)e^{-4\mu+2\mu\eta}]
		-8p_d^2(1-p_d)^2e^{-4\mu}\}. \\
	\end{split}
\end{equation}
Then we have
\begin{equation}
	\begin{aligned}
		E_Z(\ket{\Psi^-})=&[e_d Q_Z^C(\ket{\Psi^-}) + (1-e_d) Q_Z^E(\ket{\Psi^-})]/(Q_Z^C(\ket{\Psi^-})+Q_Z^E(\ket{\Psi^-})),\\
		E_X(\ket{\Psi^-})=&Q_X^E(\ket{\Psi^-})/(Q_X^C(\ket{\Psi^-})+Q_X^E(\ket{\Psi^-}).
	\end{aligned}
\end{equation}

\subsection{Charlie produces $\ket{\Psi^+}$}
In this case the correct and error gain of the Z and X bases are:
\begin{equation}
	\begin{split}
		Q_Z^C(\ket{\Psi^+})
		=&\frac{4}{N_+}\{p_d(1-p_d)^2e^{-2\mu\eta}[1-(1-p_d)e^{-2\mu\eta}]+p_d^2(1-p_d)^2e^{-4\mu}\}, \\
		Q_Z^E(\ket{\Psi^+})
		=&\frac{2}{N_+}\{(1-p_d)^2[1-(1-p_d)e^{-2\mu\eta}]^2+p_d^2(1-p_d)^2(e^{-4\mu\eta}+2e^{-4\mu})\}, \\
		Q_X^C(\ket{\Psi^+})
		=&\frac{1}{N_+}\{(1-p_d)^2[1-(1-p_d)e^{-2\mu\eta}][1-(1-3p_d)e^{-2\mu\eta}]
		+p_d^2(1-p_d)^2(e^{-4\mu\eta}+e^{-8\mu+4\mu\eta})\\
		&+(1-p_d)^2[e^{-4\mu}-(1-p_d)e^{-4\mu+2\mu\eta}][e^{-4\mu}-(1-3p_d)e^{-4\mu+2\mu\eta}]+8p_d^2(1-p_d)^2e^{-4\mu}\}, \\
		Q_X^E(\ket{\Psi^+})
		=&\frac{1}{N_+}\{(1-p_d)^2[1-(1-p_d)e^{-2\mu\eta}][1-(1-3p_d)e^{-2\mu\eta}]
		+p_d^2(1-p_d)^2(e^{-4\mu\eta}-e^{-8\mu+4\mu\eta})\\
		&-(1-p_d)^2[e^{-4\mu}-(1-p_d)e^{-4\mu+2\mu\eta}][e^{-4\mu}-(1-3p_d)e^{-4\mu+2\mu\eta}]. \\
	\end{split}
\end{equation}
Then we have
\begin{equation}
	\begin{aligned}
		E_Z(\ket{\Psi^+})=&[e_d Q_Z^C(\ket{\Psi^+}) + (1-e_d) Q_Z^E(\ket{\Psi^+})]/(Q_Z^C(\ket{\Psi^+})+Q_Z^E(\ket{\Psi^+})),\\
		E_X(\ket{\Psi^+})=&Q_X^E(\ket{\Psi^+})/(Q_X^C(\ket{\Psi^+})+Q_X^E(\ket{\Psi^+}).
	\end{aligned}
\end{equation}

We have define $E_Z$ and $E_X$ as bit error rate of the Z and X bases in ideal case (Charlie prepares $\ket{\Phi^-}$) in the letter, i.e., $E_{Z(X)}=E_{Z(X)}(\ket{\Phi^-})$.
Define $\tilde{E}_Z$ and $\tilde{E}_X$ as bit error rate of the Z and X bases in the realistic model above, we can get
\begin{equation}
	\begin{aligned}
		\tilde{E}_X=F_2 E_X+(1-F_2/3)[E_X(\Phi^+)+E_X(\Psi^-)+E_X(\Psi^+)],\\
		\tilde{E}_Z=F_2 E_Z+(1-F_2/3)[E_X(\Phi^+)+E_X(\Psi^-)+E_X(\Psi^+)].
	\end{aligned}
\end{equation}

Note that the imperfection of the entanglement source does not affect security because it can be self-generated by Eve.

We also consider the case where the cat states generated by Alice and Bob are imperfect. Since in this case the density matrix of the Z and X bases are not strictly identical, the phase error rate of the Z basis $E^{\rm ph}_{Z}$ is not equal to $E^{\rm ph}_{Z}$. To estimate the bound of $E_{\rm ph}$, we introduce
the imbalance of the quantum coin $\Delta$ defined as~\cite{lo2007security}
\begin{equation}
	\begin{aligned}
		\Delta&=\frac{1}{2Q_{Z}}[1-F(\rho_{x}^{a},\rho_{z}^{a})F(\rho_{x}^{b},\rho_{z}^{b})],\\
	\end{aligned}
\end{equation}
where $Q_Z=Q_Z^C+Q_Z^E$ is the total gain of the Z basis,
$\rho_{x}^{a(b)}$ and $\rho_{z}^{a(b)}$ are the density matrices of states in the $X$ and $Z$ bases prepared by Alice (Bob). $F(\rho_{x},\rho_{z})={\rm tr}\sqrt{\rho_{x}^{1/2}\rho_{z}\rho_{x}^{1/2}}$ is the fidelity between density matrices $\rho_{x}$ and $\rho_{z}$.
Utilizing the relationship $1-2\Delta\leq\sqrt{(1-\tilde{E}_X)(1-E^{\rm ph}_{Z})}+\sqrt{\tilde{E}_XE^{\rm ph}_{Z}}$ to estimate the bound of $E^{\rm ph}_{Z}$~\cite{lo2007security,koashi2009simple},
the secret key rate can be given by
\begin{equation}
	\begin{aligned}
		R=Q_{Z}[1-fh(\tilde{E}_Z)-h(E^{\rm ph}_{Z})].
	\end{aligned}
\end{equation}

\end{widetext}


%

\end{document}